\documentclass[a4, journal]{IEEEtran}

\usepackage{amsmath}
\usepackage{amssymb}
\usepackage{cite}
\usepackage{subfigure}
\usepackage{balance}
\usepackage{mathtools}
\usepackage{hyperref}
\hypersetup{
	unicode,
	colorlinks,
	breaklinks,
	urlcolor=cyan, 
	linkcolor=blue, 
	pdfauthor={urRehmanJ},
	pdftitle={TQE},
	pdfsubject={TQE},
	pdfkeywords={quantum noise, quantum resources, quantum states},
	pdfproducer={LaTeX},
	pdfcreator={junaid}
}

\usepackage{acronym}
\newacro{SQC}{symmetric quantum cloning}
\newacro{QKD}{quantum key distribution}
\newacro{LOCC}{local operations and classical communication}
\newacro{RL}{reinforcement learning}
\newacro{CDF}{cumulative distribution function}
\newacro{QCRLB}{quantum Cram\'er--Rao lower bound}
\newacro{QFIM}{quantum Fisher information matrix}

\newcommand{\ens}[3]{\left(#1, #2\right)_{\ket{#3}}}
\newcommand{\bO}[1]{\mathcal{O}\left(#1\right)}

\newtheorem{lemma}{Lemma}

\usepackage{graphicx, color}

\usepackage{mathrsfs} 

\usepackage{physics}
\usepackage{dsfont}

\newenvironment{proof}{\textit{Proof:}}{\hfill$\square$}

\title{
	Trading Quantum Ensembles
}
\author{Junaid ur Rehman
	\thanks{Junaid~ur~Rehman is with the Department of Electrical Engineering and the Center for Intelligent Secure Systems, King Fahd University of Petroleum and Minerals (KFUPM), Dhahran 31261, Saudi Arabia (email:junaid.urrehman@kfupm.edu.sa).}
	\thanks{Manuscript received August 18, 2025.}
}

\begin{document}
	\bstctlcite{IEEEexample:BSTcontrol}
	\maketitle
	
	\begin{abstract}
		We consider an example scenario where we require several copies of a pure quantum state $\ket{\psi}$ for some quantum information processing task. Due to practical limitations, we only have access to $N = 10^3$ depolarized copies of $\ket{\psi}$ such that the fidelity $F$ of each copy with $\ket{\psi}$ is $0.75$. We denote this quantum asset with the ensemble $\mathcal{A}: \ens{10^3}{0.75}{\psi}$. A genie appears and offers to trade $\mathcal{A}$ with either $\mathcal{B}:\ens{10^4}{0.65}{\psi}$ or with $\mathcal{C}: \ens{10^2}{0.90}{\psi}$. Should we accept the trade with either of these two ensembles? In this article, we attempt to answer this question with arbitray $N$ and $F$. More specifically, we derive resource equivalence curves from quantum resource theory of purity, quantum state distingiushibility, quantum state purification, and quantum state tomography. These curves enable ranking of these ensembles according to their operational usefulness for these tasks and allow us to answer the question of trading aforementioned ensembles. 
	\end{abstract}

	\section{Introduction}
	\label{sec:intro}
	Quantum information processing tasks require different resources, e.g., entanglement \cite{vedral_quantifying_1997, brus_characterizing_2002, horodecki_quantum_2009}, discord \cite{ollivier_quantum_2001, henderson_classical_2001, modi_classical-quantum_2012}, coherence \cite{marvian_how_2016, winter_operational_2016,  streltsov_colloquium_2017}, and purity \cite{horodecki_reversible_2003, gour_resource_2015, streltsov_maximal_2018}. The question of resourcefulness and ranking of certain states is studied under corresponding resource theories \cite{horodeckiQuantumnessContextResource2013, chitambarQuantumResourceTheories2019, sparaciariFirstLawGeneral2020, kuroiwaEveryQuantumHelps2024, kuroiwaGeneralQuantumResource2020}. These resource theories define resource monotone that characterizes the amount of resource available in a state. Then, a set of operations that are resource nonincreasing and are deemed \emph{free} is defined. This leads to the identification of maximal and free resource states. Thus, the resource theory of a certain resource is task-agnostic, providing a fundamental resource characterization.
	
	In this work, we consider a closely related but different question. Instead of task-agnostic and fandamental characterization of resources, we try capturing the resourcefulness of ensembles of noisy quantum states for certain tasks. To this end, we first consider a task, e.g., quantum state estimation, and define a relevant performance indicator, e.g., the infidelity or Bures distance. Then, we define quantum enembles to be equivalent, better, or worse than a reference ensemble if they lead to the same, better, or worse, respectively, value of the chosen performance indicator. This approach leads to analytical expressions for the contextual resource equivalence of quantum ensembles.
	
	More specifically, the main quantum state of our interest is some pure quantum state $\ket{\psi}$, but we have access to its noisy copies due to practical limitations. We denote by tuple $\ens{N}{F}{\psi}$ an ensemble of states containing $N$ independent (unentangled) depolarized copies of $\ket{\psi}$ such that the fidelity of each copy with the resource state is $F$.  That is, we have $\rho^{\otimes N}$, where each $\rho$ has the structure
	\begin{align}
		\rho = F\rho_{\psi} + \left( 1 - F\right) \rho_{\psi^{\perp}},
	\end{align}
	where $\rho_{\psi} = \ketbra{\psi}$ and $\rho_{\psi^{\perp}}$ has support orthogonal to $\ketbra{\psi}$. Using this noisy ensemble, we want to perform a task, e.g., quantum state tomography to obtain the description of $\ket{\psi}$, quantum state distinguishing, or quantum state purification. Since the main state of interest for us is a pure state, the resource theory of purity (also known as the resource theory of informational nonequilibrium) is also relevent for our discussion  \cite{horodecki_reversible_2003, gour_resource_2015}. Thus, we first find resource equivalence curves according to the resource theory of purity, which also serves as a benchmark for later developments. 
	
	\section{Main Results}
	\label{sec:MR}
	In this section we present our main analytical results. We derive resource equivalence curves of quantum ensembles for several information processing tasks.
	\subsection{Resource Theory of Informational Nonequilibrium}
	The resource theory of purity or information nonequilibrium aims at quantizing the deviation of a given quantum state from maximally mixed states \cite{streltsov_maximal_2018}. The central question of the resource theory of purity revolves around the interconvertibility of quantum states with varying levels of purity or \emph{nonuniformity} \cite{horodecki_reversible_2003, gour_resource_2015}. The asymptotic rate of distilling pure qubits from a given mixed state is
	\begin{align}
		I\left( \rho\right) = \log_2 d - S\left( \rho\right),
		\label{eq:purity_resource}
	\end{align}
	where $d$ is the dimension of $\rho$ and $S\left(\rho\right) = - \Tr{\rho\log_2 \rho}$ is the von Neumann entropy of $\rho$. We normalize \eqref{eq:purity_resource} by $\log_2 d$ to obtain the number of pure $d$-level systems that can be distilled from a given state and denote the resulting quantity by $I_d\left(\rho\right)$. 
	
	\begin{lemma}
		Given two ensembles $\mathcal{A}\colon \left(N, F\right)_{\ket{\psi}}$ and $\mathcal{B}\colon \left(M, G\right)_{\ket{\psi}}$, the ensemble $\mathcal{B}$ is more resourceful in terms of informational nonequilibrium if
		\begin{align}
			M \geq N \frac{\log_2 \left( d (d - 1)^{F - 1} \right) -h(F)}{\log_2 \left( d (d - 1)^{G - 1} \right) - h(G)},
			\label{eq:lemma_RTP}
		\end{align}
		where $h\left(x\right) = -x\log_2 x - \left(1 - x\right) \log_2 \left(1 - x\right)$ is the binary Shannon entropy of $x$.
	\end{lemma}
	\begin{proof}
	The ensemble $\mathcal{B}$ is more resourceful if we can distill more pure states by operating over it, i.e.,
	\begin{align}
		MI_d\left(\rho_B\right) \geq NI_d\left(\rho_A\right)
	\end{align}
	where $\rho_A$ and $\rho_B$ represent the density operator of a single copy of quantum state in ensemble $\mathcal{A}$ and $\mathcal{B}$, respectively. We can write the above condition more explicitly as
	\begin{align}
		M \geq N \frac{\log_2 d - S\left(\rho_A\right)}{\log_2 d - S\left(\rho_B\right)}.
	\end{align}
	Since the states in each ensemble are the depolarized copies of a pure state $\ket{\psi}$, $\rho_A$ has eigenvalues $e_0 = F$ and $e_1 = \left(1 - F\right) / \left(d - 1\right)$ with multiplicity one and $d - 1$, respectively. Similarly, $\rho_B$ has eigenvalues $e_0 = G$ and $e_1 = \left(1 - G\right) / \left(d - 1\right)$ with multiplicity one and $d - 1$, respectively. Plugging in these values for von Neumann entropy calculation and some algebraic manipulations give \eqref{eq:lemma_RTP}.
	\end{proof}
	
\begin{figure}
		\centering
		\includegraphics[width=0.475\textwidth]{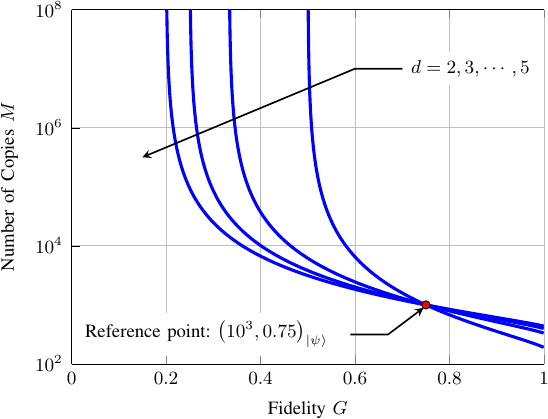}
		\caption{The resource equivalence curves based on the resource theory of informational nonequilibrium. The red dot in the center ($10^3$, $0.75$) represents the ensemble $\mathcal{A}$ and serves as the reference point. All ensembles $\ens{M}{G}{\psi}$  of dimension $d$ existing on the corresponding $d$ line are equally resourceful as the reference point, shown as red dot. Ensembles appearing above the corresponding curve are more resourceful while those appearing below are less resourceful. }
		\label{fig:rtp-crop}
	\end{figure}

	In Fig.~\ref{fig:rtp-crop}, we plot the resource equivalence curves for ${d = 2, 3, \cdots , 5}$ with respect to a reference ensemble $\ens{10^3}{0.75}{\psi}$. As expected, the $M$ increases as we decrease the fidelity of states in the ensemble with $M\to \infty$ when $G\to 1/d$. We also notice that the ensemble with a higher $d$  is more resourceful than the ensemble with a lower $d$ with same $M$ and $G$, as it leads to a higher number of $G = 1$ states.  

	\subsection{Quantum Chernoff Bound}
	The quantum Chernoff bound characterizes the behavior of probability of error in binary quantum hypothesis testing. Given $N$ copies of an unknown quantum state $\Xi \in \left\{\rho, \sigma\right\}$, we have to choose one of the two hypotheses $H_0\colon \Xi = \rho$ and $H_1\colon\Xi = \sigma$ that minimizes the probability of error, where both the hypotheses are \emph{a priori} equally likely. The behaviour of the minimum probability of error for this scenario is given by \cite{ACMB:07:PRL}
	\begin{equation}
		P_{ \mathrm{e, min}, N} \sim \exp\left(-N \xi_{\mathrm{QCB}}\right),
		\label{eq:error_probability}
	\end{equation}
	where the quantity $\xi_{\mathrm{QCB}}$ is the quantum Chernoff bound
	\begin{equation}
		\xi_{\mathrm{QCB}} = -\log\left(\min_{0 \leq s \leq 1} \tr(\rho^s \sigma^{1 - s})\right).
	\end{equation}
	
	Let us rephrase our earlier question of resource equivalence in the framework of quantum hypothesis testing. Let $\rho$ and $\sigma$ be the equally depolarized versions of two pure quantum states $\ket{r}$ and $\ket{s}$, respectively. That is
	\begin{equation}
		\rho = \lambda\left(F\right) \ketbra{r} + \left(1 - \lambda\left(F\right)\right)I/2,
	\end{equation}
	and 
	\begin{equation}
		\sigma = \lambda\left(F\right) \ketbra{s} + \left(1 - \lambda\left(F\right)\right)I/2,
	\end{equation}
	where $\lambda\left( F\right) = 2F - 1$, where we restrict the fidelity of noisy version with the pure states, i.e., $\bra{r}\rho\ket{r} = \bra{s}\sigma\ket{s} = F\in \left( 0.5, 1\right]$ so that $\lambda\left(F\right) \in \left( 0, 1 \right]$.\footnote{We have made the dependence of $\lambda$ on the fidelity $F$ explicit since the main question we ask is in terms of fidelity of the available states. We will suppress this dependence when notationally convenient. } Since the error probability of \eqref{eq:error_probability} relies on the fidelity $F$, we will often denote it as $P_{ \mathrm{e, min}, N}\left(F\right)$ Without any loss of generality we can represesnt $\ket{r}$ and $\ket{s}$ in any orthonormal basis. Let us set $\ket{r} = \ket{0}$ and $\ket{s} = \cos \frac{\theta}{2} \ket{0} + e^{\dot{\iota}\phi}\sin \frac{\theta}{2}\ket{1}$ with $\theta \in \left[0, \pi\right]$ and $\phi \in \left[0, 2\pi\right]$. where $\mathcal{B} = \left\{\ket{0}, \ket{1}\right\}$ is the orthonormal basis containing $\ket{r}$ as one of its elements. Since the state distinguishability depends on the overlap of the states, which is independent of $\phi$, there is no loss of generality in setting $\phi = 0$ for simplicity. Then, we can write
	\begin{equation}
		\rho = \frac{1 + \lambda}{2}\ketbra{0} + \frac{1 - \lambda}{2}\ketbra{1},
		\label{eq:define_rho}
	\end{equation}
	and 
	\begin{equation}
		\sigma = \frac{1 + \lambda}{2}\ketbra{s} + \frac{1 - \lambda}{2}\ketbra{s^{\perp}},
		\label{eq:define_sigma}
	\end{equation}
	where $\ket{s^{\perp}} = -\sin\frac{\theta}{2}\ket{0} + \cos \frac{\theta}{2}\ket{1}$ is the state perpendicular to $\ket{s}$. Now we can evaluate the quantity (details provided in the appendix)
	\begin{equation}
		\begin{aligned}
			\tr(\rho^s \sigma^{1 - s}) = \alpha^2 + \beta^2\left( p_+^s p_-^{1 - s} + p_-^s p_+^{1 - s}\right),
		\end{aligned}
		\label{eq:unoptimized}
	\end{equation}
	where $\alpha = \cos \frac{\theta}{2}$, $\beta = \sin \frac{\theta}{2}$, and $p_{\pm} = \frac{1 \pm \lambda}{2}$. Taking the derivative of quantity in \eqref{eq:unoptimized} with respect to $s$ and setting it equal to zero gives $s = 1/2$ as the minimizer, giving
		\begin{align}
		\xi_{\mathrm{QCB}} &= -\log \left( \alpha^2 + \beta^2\sqrt{1 - \lambda^2}\right)\\
		&= -\log\left(\alpha^2 + 2\beta^2\sqrt{F\left(1 - F\right)}\right).
			\label{eq:optimized}
		\end{align}
	Thus, we can rewrite \eqref{eq:error_probability} for our considered case as
	\begin{align}
		P_{ \mathrm{e, min}, N}\left(F\right) = \exp(N\log \left(\alpha^2  + 2\beta^2\sqrt{F\left(1 - F\right)}\right)).
	\end{align}
	Now we can address the question of resource equivalence in the context of hypothesis testing. 
	
		\begin{figure}
		\centering
		\includegraphics[width=0.48\textwidth]{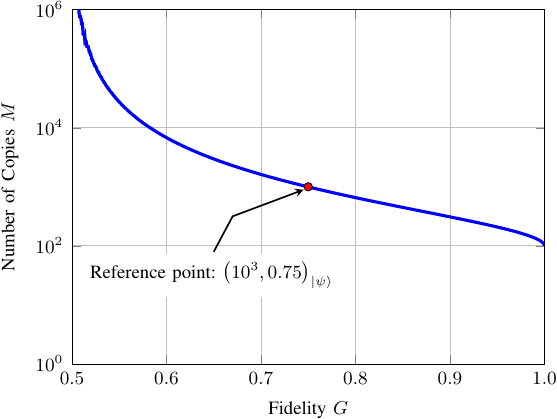}
		\caption{The resource equivalence curves based on the distinguishability of quantum states. We have fixed $\alpha = \beta = 1/\sqrt{2}$. All ensembles $\ens{M}{G}{\psi}$ exisiting on the blue line lead to the same probability of error in binary hypothese testing as the reference point, shown as red dot. Ensembles appearing above the curve offer lower error of distinguishing while those appearing below perform worse in binary state distinguishibility. }
		\label{fig:qcb-crop}
	\end{figure}

	\begin{lemma}
		Given two ensembles $\mathcal{A}\colon \left(N, F\right)_{\ket{\phi}}$ and $\mathcal{B}\colon \left(M, G\right)_{\ket{\phi}}$ with $\ket{\phi} \in \left\{\ket{r}, \ket{s}\right\}$ as described above, the ensemble $\mathcal{B}$ leads to a lower or equal probability of error in binary hypothesis testing if
		\begin{align}
			M \geq N\frac{\log \left(\alpha^2  + 2\beta^2\sqrt{F\left(1 - F\right)}\right)}{\log \left(\alpha^2  + 2\beta^2\sqrt{G\left(1 - G\right)}\right)}.
			\label{eq:main_result1}
		\end{align}
	\end{lemma}
	\begin{proof}
		We want $P_{ \mathrm{e, min}, M}\left(G\right) \leq P_{ \mathrm{e, min}, N}\left(F\right)$ or
		\begin{align}
			M\log \left(\alpha^2  + 2\beta^2\sqrt{G\left(1 - G\right)}\right) \leq N\log \left(\alpha^2  + 2\beta^2\sqrt{F\left(1 - F\right)}\right),
		\end{align}
	which leads to \eqref{eq:main_result1} since both $\log \left(\alpha^2  + 2\beta^2\sqrt{G\left(1 - G\right)}\right)$ and $\log \left(\alpha^2  +2 \beta^2\sqrt{F\left(1 - F\right)}\right)$ are negative for nondegenerate cases $\alpha \neq 1$.
	\end{proof}

	In Fig.~\ref{fig:qcb-crop}, we plot the resource equivalence curve for an ensemble with the reference point $\ens{10^3}{0.75}{\psi}$, where we have set $\alpha = \beta = 1/\sqrt{2}$.  Similar to the result of previous subsection, the resource equivalence forces $M$ to be large as the fidelity $G$ of the ensemble decreases. The number of states in the ensemble $M \to \infty$ as $G \to 1/2$.

\subsection{Purification of Quantum States}
An operational task that is related to the resource theory of purity is the purification of quantum states \cite{ciracOptimalPurificationSingle1999a,keylRateOptimalPurification2001,  yaoProtocolsTradeoffsQuantum2025, liOptimalQuantumPurity2025, childsStreamingQuantumState2025}. Given $N$ noisy copies $\rho^{\otimes N}$ of an unknown pure quantum state $\ket{\psi}$, we are interested in extracting $L < N$ higher fidelity copies of $\ket{\psi}$. Again, we restrict our initial states to be independent depolarized copies of the state of interest and set $L = 1$. In other words, the framework of purification provides a method to achieve the ensemble $\left(L = 1, H\right)_{\ket{\psi}}$ given the ensemble $\left(N>1, F\right)_{\ket{\psi}}$ with $H>F$. 

\begin{figure}[t]
	\includegraphics[width = 0.475\textwidth]{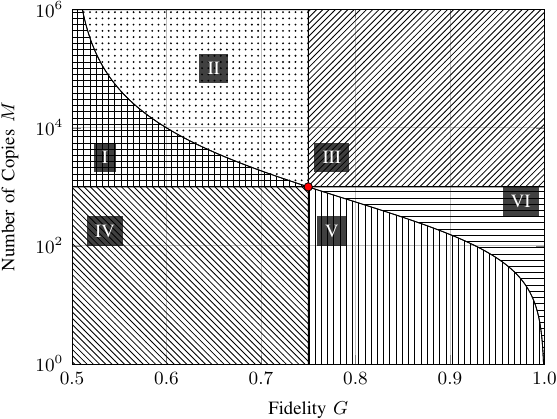}
	\caption{Resource comparison regions based on purification fidelity. The red dot in the center ($10^3$, $0.75$) represents the ensemble $\mathcal{A}$ and serves as the reference point. The separation between the regions I-II and V-VI is $M = N \left(\frac{2F - 1}{2G - 1}\right)^2\left(\frac{1 - G}{1 - F}\right)$. Loosely speaking, ensembles existing on the separation curve are equally resourceful for the task of purification whereas ensembles that exist above (resp. below) this separation curve are more (less) resourceful for the task of purification. See Lemma~\ref{TH:purification} for details.}
	\label{fig:regions}
\end{figure}

In this setting, we say two ensembles $\left(N, F\right)_{\ket{\psi}}$ and $\left(M, G\right)_{\ket{\psi}}$ are equally resourceful if they lead to the same $H$ in $\left(1, H\right)_{\ket{\psi}}$ after purification. More formally:
\begin{lemma}
	\label{TH:purification}
		Given two ensembles $\mathcal{A}\colon \left(N, F\right)_{\ket{\phi}}$ and $\mathcal{B}\colon \left(M, G\right)_{\ket{\phi}}$, the condition on $N$ such that $\mathcal{A}$ leads to a higher fidelity of purification are as follows (cf.~Fig.~\ref{fig:regions}):
		\begin{enumerate}
			\item[(A.i)] If $N \geq M$ and $F > G$ (Region~IV), $\mathcal{A}$ leads to a higher fidelity of purification.
			\item[(A.ii)] If $N \geq M$ but $F \leq G$ (Regions~V + VI), it is sufficient (but not necessary) to have $N \geq M \left(\frac{2G - 1}{2F - 1}\right)^2\left(\frac{1 - F}{1 - G}\right)$ for $\mathcal{A}$ to obtain a higher fidelity of purification (Region~V). 
			\item[(A.iii)] If $N \leq M$ but $F \geq G$ (Regions~I + II), it is necessary (but not sufficient) to have $N \geq M \left(\frac{2G - 1}{2F - 1}\right)^2\left(\frac{1 - F}{1 - G}\right)$ for $\mathcal{A}$ to obtain a higher fidelity of purification (Region~I). 
		\end{enumerate}
		Similarly, the condition on $M$ such that $\mathcal{B}$ leads to a higher fidelity of purification are as follows:
		\begin{enumerate}
			\item[(B.i)] If $M \geq N$ and $G \geq F$ (Region~III), $\mathcal{B}$ leads to a higher fidelity of purification.
			\item[(B.ii)] If $M \geq N$ but $G \leq F$ (Regions~I + II), it is sufficient (but not necessary) to have $M \geq N \left(\frac{2F - 1}{2G - 1}\right)^2\left(\frac{1 - G}{1 - F}\right)$ for $\mathcal{B}$ to obtain a higher fidelity of purification (Region~II).
			\item[(B.iii)] If $M \leq N$ but $G \geq F$ (Regions~V + VI), it is necessary but not sufficient: $M \geq N \left(\frac{2F - 1}{2G - 1}\right)^2\left(\frac{1 - G}{1 - F}\right)$ for $\mathcal{B}$ to obtain a higher fidelity of purification (Region~VI).
		\end{enumerate}
\end{lemma}
\begin{proof}
	Statements \textbf{(A.i) and (B.i)} are trivial since these conditions represent that one of the ensemble has higher number and higher quality of states in the ensemble. 
	
	For statements \textbf{(A.ii and A.iii)}, we write the expression for the fidelity of purification for $\ens{N}{F}{\psi}$ \cite[(S42)]{liOptimalQuantumPurity2025} as
	\begin{align}
		\delta_{\mathcal{A}} = \frac{1}{N}\left( 1 - \frac{1}{d}\right) \frac{1 - F}{\left( 2F - 1\right)^2} + \bO{e^{-N}},
		\label{eq:purify}
	\end{align}
	where $\delta_{\mathcal{A}} = 1 - H_{\mathcal{A}}$ is the infidelity achieved after purification of ensemble represented by $\mathcal{A}$. Writing the same for $\ens{M}{G}{\psi}$ and setting $\delta_{\mathcal{A}}\leq \delta_{\mathcal{B}}$ gives 
	\begin{align}
		N &\geq M\frac{1 - F}{1 - G}\left(\frac{2G - 1}{2F -1}\right)^2 + \bO{MNe^{-N}} - \bO{MNe^{-M}},
		\label{eq:N_condition}
	\end{align}
	where we have absorbed the factors independent of $M$ and $N$ in $\bO{\cdot}$. Noting that the correction term $\bO{MNe^{-N}} - \bO{MNe^{-M}}$ is negative (resp. positive) when $N > M$(resp. $N < M$), i.e., as in the statement A.ii (resp. A.iii), it is sufficient but not necessary (resp. necessary but not sufficient) to have $N \geq M \left(\frac{2G - 1}{2F - 1}\right)^2\left(\frac{1 - F}{1 - G}\right)$ to satisfy \eqref{eq:N_condition}.
	
	Statements \textbf{(B.ii) and (B.iii)} can also be proven by similar reasoning after setting $\delta_{\mathcal{B}}\leq \delta_{\mathcal{A}}$.
\end{proof}

We summarize this result as follows. $\mathcal{B}$ is more resourceful in purification if it exists in region II or III. Similarly, $\mathcal{B}$ is less resourceful if it exists in region IV or V. Furthermore, for $G\geq F$ (resp. $F\geq G$), it is necessary for $\mathcal{B}$ to exist above (resp. below) the separation line to be more (less) resourceful than $\mathcal{A}$.

\subsection{Quantum State Estimation}
The final operational task that we consider here is the quantum state tomography or quantum state estimation. Quantum state tomography attempts to provide as accurate as possible classical description of a quantum state by measuring several of its copies. The quality of the classical description is given by some distance measure with the actual state, i.e., the ground truth. Similar to the previous operationals tasks, we assume the ground truth state to be a pure state $\ket{\psi}$. Ideally, a large number of its copies should be available for its estimation. Instead, we have access to $\ens{N}{F}{\psi}$. 
The performance of quantum state estimation is bounded by the \ac{QCRLB} for the estimation of its parameters and by the Gill-Massar bound for its performance measure, e.g., for infidelity or the Bures distance. 

Let us begin by calculating the \ac{QCRLB} for this specific scenario from where the Gill-Massar like inequality can be derived. The \ac{QFIM} is defined as \cite{liuQuantumFisherInformation2019}
\begin{align}
	\left[\mathcal{F}\right]_{a, b} = \frac{1}{2}\tr(\rho \left\{L_a, L_b\right\}), \quad a, b \in \left\{\theta, \phi\right\},
\end{align}
 where $\left\{\cdot, \cdot\right\}$ is the anticommutator and $L_a$ is the symmetric logarithmic derivative of parameter $a$, implicitly defined via the Lyapunov equation
 \begin{align}
 	\derivative{\rho}{a} = \frac{\rho L_a + L_a\rho}{2}.
 \end{align}
Let us assume the Bloch parameterization of our state of interest, i.e., $\ket{\psi} = \cos \frac{\theta}{2} \ket{0} + e^{\iota \phi}\sin \frac{\theta}{2}\ket{1}$.  We can explicitly calculate the \ac{QFIM} of the depolarized state as
\begin{align}
	\mathcal{F} = \begin{bmatrix}
		\lambda^2 & 0 \\ 0 & \lambda^2 \sin^2(\theta)
	\end{bmatrix}.
\end{align}
The \ac{QCRLB} is
\begin{align}
	\mathcal{C} \geq \frac{1}{N}{\mathcal{F}^{-1}},
\end{align}
where $\mathcal{C}$ denotes the covariance matrix of estimated parameters $\hat{\theta}$ and $\hat{\phi}$. We can obtain the Gill-Massar like inequality
\begin{align}
	\mathds{E}\left[D_{\mathrm{B}}^2\right] &= \tr(g_{\mathrm{pure}} \mathcal{C})\label{eq:mean_squared_bures}\\
	 &\geq \frac{1}{N}\tr(g_{\mathrm{pure}} \mathcal{F}^{-1})\\
		&= \frac{1}{4N\left( 2F - 1\right)^2},\label{eq:GMB}
\end{align}
where $D_{\mathrm{B}}$ is the Bures distance, $\mathds{E}\left[\cdot \right]$ is the expectation over the set of density operators, and $g_{\mathrm{pure}} = \frac{1}{4\lambda^2} \mathcal{F}$ is the Bures metric for \emph{pure} state parameterized by Bloch angles as above \cite{paris_quantum_2009}.\footnote{The Bures metric is defined as $g = \frac{1}{4}\mathcal{F}$. Since our target state is the pure state, we use the Bures metric for pure states $g_{\mathrm{pure}} = \frac{1}{4}\mathcal{F}_{\mathrm{pure}}$, where $\mathcal{F}_{\mathrm{pure}} = \frac{1}{\lambda^2}\mathcal{F}$ is the \ac{QFIM} for the pure state. However, we use the \ac{QFIM} for depolarized state since the available copies are depolarized. } The inequality above can be saturated by collective measurements over $N$ copies of the available state followed by appropriate post-processing.

	\begin{lemma}
	Given two ensembles $\mathcal{A}\colon \left(N, F\right)_{\ket{\psi}}$ and $\mathcal{B}\colon \left(M, G\right)_{\ket{\psi}}$ the ensemble $\mathcal{B}$ leads to a lower or equal mean error of quantum state estimation iff
	\begin{align}
		M \geq N\left(\frac{2F - 1}{2G - 1}\right)^2.
		\label{eq:main_result3}
	\end{align}
	\end{lemma}
	\begin{proof}
		Let $d_{\mathcal{A}} = 1/4N\left(2F - 1\right)^2$ and $d_{\mathcal{B}} = 1/4M\left(2G - 1\right)^2$ be the mean squared Bures distances between the estimated and the actual state achievable with $\mathcal{A}$ and $\mathcal{B}$, respectively, as given in \eqref{eq:mean_squared_bures}-\eqref{eq:GMB}. Setting $d_{\mathcal{B}} \leq d_{\mathcal{A}}$ gives \eqref{eq:main_result3}. 
	\end{proof}

\begin{figure}
	\centering
	\includegraphics[width=0.48\textwidth]{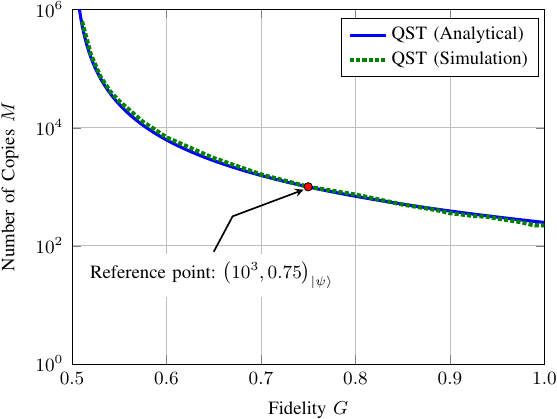}
	\caption{The theoretical and simulations based resource equivalence curves based on the quantum state tomography. All ensembles $\ens{M}{G}{\psi}$ exisiting on the blue line lead to the same error (Bures distance) of quantum state estimation as the reference point, shown as red dot. Ensembles appearing above the curve offer lower error of estimation while those appearing below perform worse in the quantum state estimation. }
	\label{fig:qst-crop}
\end{figure}

	In Fig.~\ref{fig:qst-crop}, we plot the resource equivalence curve for the task of quantum state estimation. All ensembles existing on the resource equivalence curve leads to the same Bures distance of estimation as the reference ensemble $\ens{10^3}{0.75}{\psi}$. Here, in addition to the analytical resource equivalence derived above, we also plot the simulation based resource equivalence curve. To this end, we performed standard quantum state tomography of Haar random state over a grid of $\left(N, F\right)$ space for $10^3$ random states at each data point. Since we measured depolarized copies of $\ket{\psi}$, we performed simple quantum error mitigation by picking the largest eigenvector of constructed state as the estimate $\ket{\hat{\psi}}$ of $\ket{\psi}$. This does not require the knowledge of $F$ or $\ket{\psi}$. We plotted the resulting infidelity values as a contour plot and adjusted the contours such that one of them passes through our reference point, shown in red. The coordintates of that contour were extracted and are being plotted in Fig.~\ref{fig:qst-crop} as the simulation result.

\section{Comparison of Results and Conclusion}

\begin{figure}
	\centering
	\includegraphics[width=0.48\textwidth]{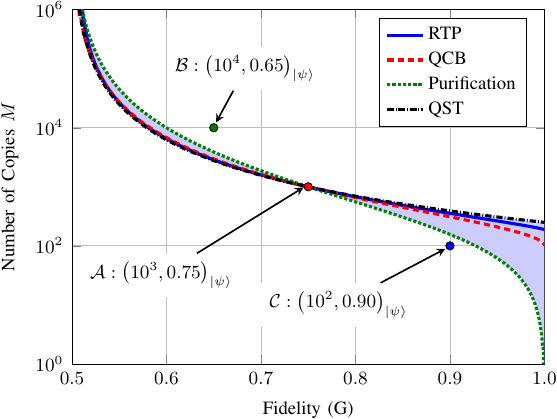}
	\caption{Comparison of all resource equivalence curves. The resource equivalence curves resulting from the resource theory of purity (RTP), quantum Chernoff bound (QCB), and quantum state tomography (QST) behave very similarly. The curve originating from the purification has similar trend but the numerical values differ significantly from the other resource equivalence curves. Thus, the ensembles $\ens{M}{G}{\psi}$ existing in the shaded blue region may be better or worse than the reference point depending on the operational task. All the ensembles appearing above the shaded region are better than the reference while those appearing below are worse. Ensembles $\mathcal{A}$, $\mathcal{B}$, and $\mathcal{C}$ are the ones mentioned in the abstract. }
	\label{fig:comp}
\end{figure}

In this section we compare various resource equivalence curves that we obtained in the previous section. In Fig.~\ref{fig:comp}, we plot all resource equivalence curves ($d = 2$, where applicable). We observe that three resource equivalence curves (resource theory of purity (RTP), quantum Chernoff bound (QCB), and quantum state tomography (QST)) behave very similarly. However, the resource equivalence curve of purification behaves differently. In particular, the reference ensemble $\ens{10^3}{0.75}{\psi}$ is equivalent to ensembles of $10^2$ to $10^{2.3}$ fidelity-1 particles for QCB, RTP, and QST. However, the same reference ensemble is equivalent to $\approx 1$ fidelity-1 particle for the task of purification. Similarly, for ensembles with lower fidelity than the reference ensemble, the purification resource equivalence result in a higher number of fidelity $G$ particles as compared to RTP, QCB, and QST. Thus, the purification resource equivalence has a more steep slope as compared to the other resource equivalence curves. We can summarize this behavior as: the purification-based resource equivalence sacrifices more copies for the gain of fidelity.

The differences in the resource equivalence curves can potentially be explained by the setting of each considered task. For example, in the resource theory of purity or informational nonequilibrium, the allowable operations are limited to the certain free operations defined in this resource theory \cite{horodecki_reversible_2003, gour_resource_2015}. No such restriction on the allowable set of operations exist in the other resource equivalence curves that we considered here. A key difference among three operational tasks is the output of the protocol. For state tomography the output is the classical description of $\ket{\psi}$, for distinguishibility the output is the accepted hypothesis, which is again classical. In contrast, the output of purification protocol is a quantum state. This can be a potential reason for the distinct resource equivalence curve for purification. Another possible contributing factor in these differences can be the neglected higher order terms in deriving the base results that served as the starting point for each case in this manuscript. 

In Fig.~\ref{fig:comp}, we also mark the region (shaded blue) where the resource equivalence is task specific. Ensembles existing in this region can be more or less useful than the reference ensemble depending on the task. All ensembles existing outside of this shaded region have unambiguous ranking as compared to the reference ensemble, with respect to resource equivalence curves discussed in this article. We also mark the locations of the ensembles mentioned in the abstract of this article. The direct answer to the question posed about trading our resources is: we should accept the trade of $\mathcal{A}$ with $\mathcal{B}$ because it exists above the resource equivalence curves for $\mathcal{A}$ for all the curves. 

The framework developed in this manuscript can also be applied to benchmark and rank quantum information sources. In this context, the ensemble size corresponds to the source’s generation rate, while noisy state preparation is considered free---assuming the generator is readily available. However, the process remains constrained by both the generation rate and the fidelity of the produced states. For operational tasks such as quantum state estimation, one can reach any target estimation error by trading off sampling duration. Under this formulation, the same framework introduced earlier can be employed to evaluate and rank different quantum state generators.

Future works may consider more operational tasks and attempt to identify, in more detail, the reasons for subtle differences between the resource equivalence curves.


\appendix
\section*{Deriving \eqref{eq:unoptimized} and \eqref{eq:optimized}}
Using \eqref{eq:define_rho} and \eqref{eq:define_sigma}, we can write
	\begin{align}
	\tr(\rho^s\sigma^{1 - s}) &= \mathrm{tr}\left(\left(p_+^s\ketbra{0} + p_-^s\ketbra{1}\right)\right.\nonumber\\
	&~~~~~~~~~\left.\left(p_+^{1 - s}\ketbra{s} + p_-^{1 - s}\ketbra{s^\perp}\right)\right)\\
	&= \mathrm{tr}\left( p_+\alpha\ketbra{0}{s} - p_+^sp_-^{1 - s}\beta\ketbra{0}{s^{\perp}} + \right.\nonumber\\
	&~~~~~~~~~\left.
	p_-^sp_+^{1 - s}\beta\ketbra{1}{s} + p_-\alpha\ketbra{1}{s^{\perp}}
	\right),
	 \end{align}
	 where the second equality follows from the facts $\braket{0|s} = \braket{1|s^{\perp}} = \alpha$ and $\braket{1|s} = -\braket{0|s^{\perp}} = \beta$. Using the linearity and the cyclic property of trace, we obtain \eqref{eq:unoptimized}, i.e., 
	 \begin{equation*}
	 	\begin{aligned}
	 		\tr(\rho^s \sigma^{1 - s}) = \alpha^2 + \beta^2\left( p_+^s p_-^{1 - s} + p_-^s p_+^{1 - s}\right).
	 	\end{aligned}
	 \end{equation*}
	 Next, we differentiate both sides with respect to $s$
	 \begin{align}
	 	\derivative{}{s}\tr(\rho^s \sigma^{1 - s}) &= \beta^2\derivative{ }{s}\left(p_+^s p_-^{1 - s} + p_-^s p_+^{1 - s}\right).
	 \end{align}
	 Setting it equal to zero, we can drop $\beta$ and obtain
	 \begin{align}
	 	&\derivative{ }{s}\left(p_+^s p_-^{1 - s} + p_-^s p_+^{1 - s}\right)\\
	 	&= \log p_+ - \log p_- + p_+^{1 - 2s}p_-^{2s - 1}\log p_- - \nonumber \\
	 	&~~~~~~~~~~~~~~~~~~~ p_+^{1 - 2s}p_-^{2s - 1}\log p_+\\
	 	&= \log p_+\left( 1 - p_+^{1 - 2s}p_-^{2s - 1}\right) - \nonumber \\
	 	&~~~~~~~~~~~~~~~~~~~\log p_-\left( 1 - p_+^{1 - 2s}p_-^{2s - 1}\right)\\
	 	&= \left(\log p_+ - \log p_-\right) \left( 1 - p_+^{1 - 2s}p_-^{2s - 1}\right) = 0.
	 \end{align}
	 This gives two possible solutions, one of which is degenerate, i.e., $s \in \left[ 0, 1\right]$ when $p = 0$, and $s = 1/2$ for any $p \in \left[ 0, 1\right]$. The expression \eqref{eq:optimized} can be obtained by plugging in $s = 1/2$ in \eqref{eq:unoptimized} and using $\lambda = 2F - 1$.
	\balance

\end{document}